\documentclass{PoS-hep}
\usepackage{amsmath}
\usepackage{bm}
\usepackage{epsfig}
\usepackage{graphics}
\usepackage{upgreek}
\usepackage{amsfonts}
\usepackage{amsbsy}
\usepackage{amscd}
\usepackage{bbm}
\usepackage{eufrak}
\usepackage{cite}

\renewenvironment{subequations}{%
\refstepcounter{equation}%
\setcounter{parentequation}{\value{equation}}%
  \setcounter{equation}{0}
  \ignorespaces
}{%
  \setcounter{equation}{\value{parentequation}}%
  \ignorespacesafterend
}
\newcommand{\eeq}{\end{equation}}
\newcommand{\beq}{\begin{equation}}
\newcommand{\ba}{\begin{array}}
\newcommand{\ea}{\end{array}}
\newcommand{\bea}{\begin{eqnarray}}
\newcommand{\eea}{\end{eqnarray}}
\newcommand{\baq}{\begin{eqnarray}}
\newcommand{\eaq}{\end{eqnarray}}

\newcommand{\beqs}{\begin{subequations}}
\newcommand{\eeqs}{\end{subequations}}
\newcommand{\eec}{\end{center}}
\newcommand{\bec}{\begin{center}}
\newcommand{\eem}{\end{matrix}}
\newcommand{\bem}{\begin{matrix}}
\newcommand{\Eref}[1]{Eq.~(\ref{#1})}
\newcommand{\Sref}[1]{Sec.~\ref{#1}}
\newcommand{\Fref}[1]{Fig.~\ref{#1}}
\newcommand{\Tref}[1]{Table~\ref{#1}}
\newcommand{\cref}[1]{Ref.~\cite{#1}}
\newcommand{\crefs}[1]{Refs.~\cite{#1}}
\newcommand{\etal}{{\it et al.\/}}

\newcommand\eqs[2]{Eqs.~(\ref{#1}) and (\ref{#2})}

\newcommand{\sEref}[2]{Eq.~(\ref{#1}{\ftn\sf {#2}})}
\newcommand{\sFref}[2]{Fig.~\ref{#1}-{\ftn\sf ({#2})}}

\newcommand{\ftn}{\footnotesize}

\newcommand{\TeV}{{\mbox{\rm TeV}}}

\newcommand{\GeV}{{\mbox{\rm GeV}}}
\newcommand{\eV}{{\mbox{\rm eV}}}

\def\to{\rightarrow}

\def\llgm{\left\lgroup}
\def\rrgm{\right\rgroup}
\def\lf{\left(}
\def\rg{\right)}
\newcommand\vev[1]{\langle {#1} \rangle}
\newcommand{\Gr}{\ensuremath{\widetilde{G}}}
\newcommand{\Yb}{\ensuremath{Y_{B}}}
\newcommand{\Yg}{\ensuremath{Y_{3/2}}}

\newcommand{\Vhi}{\ensuremath{\widehat V_{\rm HI}}}
\newcommand{\Hhi}{\ensuremath{\widehat H_{\rm HI}}}

\newcommand{\Khi}{\ensuremath{K}}

\newcommand{\mP}{\ensuremath{m_{\rm P}}}

\newcommand{\Mgut}{\ensuremath{M_{\rm GUT}}}
\newcommand{\Qef}{\ensuremath{\Lambda_{\rm UV}}}
\newcommand{\Ggut}{\ensuremath{G_{B-L}}}

\newcommand{\lm}{\ensuremath{\lambda_\mu}}

\def\openone{\leavevmode\hbox{\small1\kern-3.8pt\normalsize1}}

\newcommand{\Gsn}{\ensuremath{\what{\Gamma}_{\rm \dph}}}
\newcommand{\GNsn}{\ensuremath{\what{\Gamma}_{\dph\to N_i^cN_i^c}}}
\newcommand{\Ghsn}{\ensuremath{\what{\Gamma}_{\dph\to \hu\hd}}}

\newcommand{\msn}{\ensuremath{\what m_{\rm \dph}}}

\newcommand{\hd}{{\ensuremath{H_d}}}
\newcommand{\hu}{{\ensuremath{H_u}}}

\newcommand{\dphi}{\ensuremath{\what{\delta\phi}}}
\newcommand{\dph}{\ensuremath{\delta\phi}}

\newcommand{\ks}{\ensuremath{k_\star}}
\newcommand{\Ns}{\ensuremath{{\what N_\star}}}
\newcommand{\ns}{\ensuremath{n_{\rm s}}}

\newcommand{\nb}{\ensuremath{N_{X}}}
\newcommand{\as}{\ensuremath{a_{\rm s}}}
\newcommand{\As}{\ensuremath{A_{\rm s}}}
\newcommand{\rw}{\ensuremath{r_{0.002}}}
\newcommand{\rs}{\ensuremath{r_{\pm}}}
\newcommand{\rpm}{\ensuremath{r_{\pm}}}

\newcommand{\Ve}{\ensuremath{\widehat{V}}}

\newcommand{\sni}{\ensuremath{N^c_i}}
\newcommand{\ssni}{\ensuremath{\widetilde N_i^c}}

\newcommand{\aS}{\ensuremath{{\rm a}_S}}
\newcommand{\Ald}{\ensuremath{A_\lambda}}
\newcommand{\am}{\ensuremath{{\rm a}_{3/2}}}

\newcommand{\mrh[1]}{\ensuremath{M_{#1N^c}}}
\newcommand{\lrh[1]}{\ensuremath{\lambda_{#1N^c}}}

\newcommand{\mD[1]}{\ensuremath{m_{#1\rm D}}}
\newcommand{\mn[1]}{\ensuremath{m_{#1\nu}}}

\newcommand{\wrhn[1]}{\ensuremath{N^c_{#1}}}

\newcommand{\Whi}{\ensuremath{W_{\rm HI}}}

\def\ve{\varepsilon}

\def\bbet{{\bar\beta}}
\def\al{{\alpha}}

\def\n{\bar{n}}

\def\th{{\theta}}
\def\thb{{\bar\theta}}
\def\thn{{\theta_{\Phi}}}

\newcommand{\Trh}{\ensuremath{T_{\rm rh}}}
\newcommand{\sg}{\ensuremath{\phi}}

\newcommand{\ld}{\ensuremath{\lambda}}
\newcommand{\ldu}{\ensuremath{\uplambda}}

\newcommand{\kp}{\ensuremath{\kappa}}

\newcommand{\sgx}{\ensuremath{\phi_\star}}
\newcommand{\sgf}{\ensuremath{\phi_{\rm f}}}

\newcommand{\what}{\ensuremath{\widehat}}

\newcommand{\se}{\ensuremath{\widehat \phi}}
\newcommand{\sex}{\ensuremath{\widehat{\phi}_\star}}

\newcommand{\eph}{\ensuremath{\widehat \epsilon}}

\newcommand{\mgr}{\ensuremath{m_{3/2}}}
\newcommand{\mg}{{\ensuremath{M_{1/2}}}}

\def\Kap{K\"{a}hler potential}

\def\sub{subplanckian}

\def\bcp{{\sc\small Bicep2}/{\it Keck Array}}
\newcommand{\plk}{{\it Planck}}

\newcommand{\diag}{\ensuremath{{\sf diag}}}
\newcommand{\im}{\ensuremath{{\sf Im}}}

\newcommand{\cm}{\ensuremath{c_{-}}}
\newcommand{\cp}{\ensuremath{c_{+}}}
\newcommand{\fm}{\ensuremath{F_{-}}}
\newcommand{\fp}{\ensuremath{F_{+}}}

\newcommand{\nsu}{\ensuremath{{N_X}}}

\newcommand{\Gbl}{\ensuremath{G_{B-L}}}

\newcommand{\fr}{\ensuremath{f_{R}}}
\newcommand{\frs}{\ensuremath{f_{R\star}}}

\newcommand{\fns}{\ensuremath{f_{R\star}}}

\newcommand{\phc}{\ensuremath{\Phi}}
\newcommand{\phcb}{\ensuremath{\bar\Phi}}
\newcommand\mtta[4]{\mbox{
$\llgm\bem #1 &#2 \cr #3& #4\eem\rrgm$}}

\newcommand{\bdhh}{{\ensuremath{\normalsize I{\kern-2.9pt H}}}}

\title{\boldmath $B-L$ Higgs Inflation in Supergravity With Several Consequences}

\ShortTitle{B-L HI in SUGRA With Several Consequences}

\author{\speaker{Constantinos Pallis}\\
       Department of Physics, University of Cyprus\\
       E-mail: \email{kpallis@gen.auth.gr}}

\abstract{We consider a renormalizable extension of the minimal
supersymmetric standard model endowed by an R and a gauged $B - L$
symmetry. The model incorporates chaotic inflation driven by a
quartic potential, associated with the Higgs field which leads to
a spontaneous breaking of $U(1)_{B-L}$, and yields possibly
detectable gravitational waves. We employ semi-logarithmic Kahler
potentials with an enhanced shift symmetry which include only
quadratic terms and integer prefactors for the logarithms. An
explanation of the $\mu$ term of the MSSM is also provided,
consistently with the low energy phenomenology, under the
condition that a related parameter in the superpotential is
somewhat small. Baryogenesis occurs via non-thermal leptogenesis
which is realized by the inflaton's decay to the lightest and/or
next-to-lightest right-handed neutrinos.}

\FullConference{Corfu Summer Institute 2017 "School and Workshops on Elementary Particle Physics and Gravity"\\
                 2-28 September 2017\\
                 Corfu, Greece}

\begin{document}

\section{Introduction}

We concentrate on the theoretically most promising models of
kinetically modified non-minimal \emph{Higgs inflation} ({\sf\ftn
HI}) investigated in \cref{univ}, considering exclusively integer
prefactors for the logarithms included in the \Kap s. We embed the
selected models in a complete framework which presented in
Sec.~\ref{fhim}. The inflationary part of this context is
described in Sec.~\ref{fhi}.  Then, in Sec.~\ref{secmu}, we
explain how the \emph{minimal supersymmetric standard model}
({\sf\ftn MSSM}) is obtained as low energy theory and, in
Sec.~\ref{pfhi}, we outline how the observed \emph{baryon
asymmetry of the universe} ({\ftn\sf BAU}) is generated via
\emph{non-thermal leptogenesis} ({\sf\ftn nTL}). Our conclusions
are summarized in Sec.~\ref{con}. Throughout the text, the
subscript of type $,z$ denotes derivation \emph{with respect to}
(w.r.t) the field $z$ and charge conjugation is denoted by a star.
Unless otherwise stated, we use units where $\mP = 2.433\cdot
10^{18}~\GeV$ is taken unity.

\section{Model Description}\label{fhim}

We focus on a ``\emph{Grand Unified Theory}'' ({\sf \ftn GUT})
based on $\Ggut=G_{\rm SM}\times U(1)_{B-L}$, where ${G_{\rm SM}}=
SU(3)_{\rm C}\times SU(2)_{\rm L}\times U(1)_{Y}$ is the gauge
group of the standard model and $B$ and $L$ denote the baryon and
lepton number respectively. The superpotential of our model
naturally splits into two parts:
\beq W=W_{\rm MSSM}+\Whi,\>\>\>\mbox{where}\label{Wtot}\eeq
\paragraph{\sf\ftn (a)} $W_{\rm MSSM}$ is the part of $W$ which contains the
usual terms -- except for the $\mu$ term -- of MSSM, supplemented
by Yukawa interactions among the left-handed leptons ($L_i$) and
$\sni$:
\beqs \beq W_{\rm MSSM} = h_{ijD} {d}^c_i {Q}_j \hd + h_{ijU}
{u}^c_i {Q}_j \hu+h_{ijE} {e}^c_i {L}_j \hd+ h_{ijN} \sni L_j \hu.
\label{wmssm}\eeq
Here the $i$th generation $SU(2)_{\rm L}$ doublet left-handed
quark and lepton superfields are denoted by $Q_i$ and $L_i$
respectively, whereas the $SU(2)_{\rm L}$ singlet antiquark
[antilepton] superfields by $u^c_i$ and ${d_i}^c$ [$e^c_i$ and
$\sni$] respectively. The electroweak Higgs superfields which
couple to the up [down] quark superfields are denoted by $\hu$
[$\hd$].

\paragraph{\sf\ftn (b)} $\Whi$ is the part of $W$ which is relevant for
HI, the generation of the $\mu$ term of MSSM and the Majorana
masses for $\sni$'s. It takes the form
\beq\label{Whi} \Whi= \ld S\lf \bar\Phi\Phi-M^2/4\rg+\lm
S\hu\hd+\lrh[i]\phcb N^{c2}_i\,. \eeq\eeqs
The imposed $U(1)_R$ symmetry ensures the linearity of $\Whi$
w.r.t $S$. This fact allows us to isolate easily via its
derivative the contribution of the inflaton into the F-term SUGRA
potential, placing $S$ at the origin -- see \Sref{fhi1}. It plays
also a key role in the resolution of the $\mu$ problem of MSSM via
the second term in the \emph{right-hand side} ({\sf\ftn r.h.s}) of
\Eref{Whi} -- see \Sref{secmu2}. The inflaton is contained in the
system $\bar\Phi - \Phi$. We are obliged to restrict ourselves to
\sub\ values of $\bar\Phi\Phi$ since the imposed symmetries do not
forbid non-renormalizable terms of the form $(\bar\Phi\Phi)^{p}$
with $p>1$ -- see \Sref{fhi2}. The third term in the r.h.s of
\Eref{Whi} provides the Majorana masses for the $\sni$'s  and
assures the decay of the inflaton to $\ssni$, whose subsequent
decay can activate nTL. Here, we work in the so-called
\emph{$\sni$-basis}, where $\mrh[i]$ is diagonal, real and
positive. These masses, together with the Dirac neutrino masses in
Eq.~(\ref{wmssm}), lead to the light neutrino masses via the
seesaw mechanism.

\renewcommand{\arraystretch}{1.1}

\begin{table}[!t]
\begin{center}
\begin{tabular}{|c|c|c|c|c|}\hline
{\sc Superfields}&{\sc Representations}&\multicolumn{3}{|c|}{\sc
Global Symmetries}\\\cline{3-5}
&{\sc under $\Gbl$}& {\hspace*{0.3cm} $R$\hspace*{0.3cm} }
&{\hspace*{0.3cm}$B$\hspace*{0.3cm}}&{$L$} \\\hline\hline
\multicolumn{5}{|c|}{\sc Matter Fields}\\\hline
{$e^c_i$} &{$({\bf 1, 1}, 1, 1)$}& $1$&$0$ & $-1$ \\
{$N^c_i$} &{$({\bf 1, 1}, 0, 1)$}& $1$ &$0$ & $-1$
 \\
{$L_i$} & {$({\bf 1, 2}, -1/2, -1)$} &$1$&{$0$}&{$1$}
\\
{$u^c_i$} &{$({\bf 3, 1}, -2/3, -1/3)$}& $1$  &$-1/3$& $0$
\\
{$d^c_i$} &{$({\bf 3, 1}, 1/3, -1/3)$}& $1$ &$-1/3$& $0$
 \\
{$Q_i$} & {$({\bf \bar 3, 2}, 1/6 ,1/3)$} &$1$ &$1/3$&{$0$}
\\ \hline
\multicolumn{5}{|c|}{\sc Higgs Fields}\\\hline
{$\hd$}&$({\bf 1, 2}, -1/2, 0)$& {$0$}&{$0$}&{$0$}\\
{$\hu$} &{$({\bf 1, 2}, 1/2, 0)$}& {$0$} & {$0$}&{$0$}\\
\hline
{$S$} & {$({\bf 1, 1}, 0, 0)$}&$2$ &$0$&$0$  \\
{$\Phi$} &{$({\bf 1, 1}, 0, 2)$}&{$0$} & {$0$}&{$-2$}\\
{$\bar \Phi$}&$({\bf 1, 1}, 0,-2)$&{$0$}&{$0$}&{$2$}\\
\hline\end{tabular}
\end{center}
\caption[]{\sl \small The representations under $\Gbl$ and the
extra global charges of the superfields of our model.}\label{tab1}
\end{table}
\renewcommand{\arraystretch}{1.}

HI is feasible if $\Whi$ cooperates with \emph{one} of the
following \Kap s:
\beqs\bea
K_1&=-3\ln\left(1+\cp\fp+F_{1X}(|X|^2)\right)+\cm\fm&~~\mbox{with}~~F_{1X}=-\ln\left(1+|X|^2/N\right),\label{K1}\\
K_2&=-2\ln\left(1+\cp\fp\right)+\cm\fm+F_{2X}(|X|^2)&~~\mbox{with}~~F_{2X}=N_X\ln\left(1+|X|^2/N_X\right),\label{K2}\\
K_3&=-2\ln\left(1+\cp\fp\right)+F_{3X}(\fm,
|X|^2)~~~~~~~&~~\mbox{with}~~F_{3X}=N_X\ln\left(1+|X|^2/N_X+\cm\fm/N_X\right),~~
\label{K3} \eea\eeqs
where $F_\pm=\left|\Phi\pm\bar\Phi^*\right|^2$, $0<N_X<6$,
$X^\gamma=S,\hu,\hd,\ssni$  and the complex scalar components of
the superfields $\Phi, \bar\Phi, S, \hu$ and $\hd$ are denoted by
the same symbol whereas this of $\sni$ by $\ssni$. The functions
$F_\pm$ assist us in the introduction of a shift symmetry for the
Higgs fields  -- cf. \cref{jhep}. In all $K$'s, $\fp$ is included
in the argument of a logarithm with coefficient $(-3)$ or $(-2)$
whereas $\fm$ is outside it -- cf. \cref{nMHkin}. As regards the
non-inflaton fields $X^\gamma$, we assume that they have identical
kinetic terms expressed by the functions $F_{lX}$ with $l=1,2,3$
and their form is given in \cref{univ}. Just for definiteness, we
here adopt the logarithmic forms. These functions ensures the
stability and the heaviness and of these modes \cite{su11}
including \emph{exclusively} quadratic terms. In the limits
$\cp\to0$ and $\ld\to0$, our models are completely natural in the
't Hooft sense, since they enjoy the following enhanced symmetries
\beq \Phi \to\ \Phi+c,\>\>\>\bar\Phi \to\ \bar\Phi+c^*
\>\>\>\mbox{and}\>\>\> X^\gamma \to\ e^{i\varphi_\gamma}
X^\gamma,\label{shift}\eeq
where $c$ and $\varphi_\gamma$ are complex and real numbers
respectively and no summation is applied over $\gamma$.

\section{Inflationary Scenario}\label{fhi}

The salient features of our inflationary scenario are studied at
tree level in \Sref{fhi1} and at one-loop level in \Sref{fhi11}.
We then present its predictions in \Sref{fhi3}, calculating a
number of observable quantities introduced in Sec.~\ref{fhi2}.

\subsection{Inflationary Potential}\label{fhi1}

If we express $\Phi, \bar\Phi$ and $X^\gamma= S,\hu,\hd,\ssni$
according to the parametrization
\beq\label{hpar} \Phi=\frac{\sg e^{i\th}}{\sqrt{2}}
\cos\thn,~~\bar\Phi=\frac{\sg e^{i\thb}}{\sqrt{2}}
\sin\thn~~\mbox{and}~~X^\gamma= \frac{x^\gamma +i\bar
x^\gamma}{\sqrt{2}}\,,\eeq
where $0\leq\thn\leq\pi/2$, we can easily deduce that the
\emph{Einstein frame} SUGRA scalar potential $\Ve$ which can be
found via the formula
\beq \Ve=\Ve_{\rm F}+ \Ve_{\rm D}\>\>\>\mbox{with}\>\>\> \Ve_{\rm
F}=e^{\Khi}\left(K^{\al\bbet}D_\al \Whi D^*_\bbet \Whi^*-3{\vert
\Whi\vert^2}\right) \>\>\>\mbox{and}\>\>\>\Ve_{\rm D}=
{1\over2}g^2 \sum_{\rm a} {\rm D}_{\rm a} {\rm D}_{\rm
a},\label{Vsugra} \eeq
exhibit a D-flat direction at
\beq \label{inftr} x^\gamma=\bar
x^\gamma=\th=\thb=0\>\>\>\mbox{and}\>\>\>\thn={\pi/4}\,.\eeq
Along this the only surviving term of $\Ve$ can be written
universally as
\beq \label{Vhi} \Vhi= e^{K}K^{SS^*}\, |W_{{\rm
HI},S}|^2=\frac{\ld^2(\sg^2-M^2)^2}{16\fr^{2}}\>\>\>\mbox{where}\>\>\>\fr=1+\cp\sg^2\eeq
plays the role of a non-minimal coupling to Ricci scalar in the
\emph{Jordan frame} -- see \cref{jhep}. Clearly $\Vhi$ develops an
inflationary plateau as in the original case of non-minimal
inflation \cite{plin}. Contrary to that case, though, here we have
also $\cm$ which dominates the canonical normalization of $\phi$
and allows for distinctively different inflationary outputs as
shown in \crefs{jhep, nMHkin}. To specify it together with the
normalization of the other fields, we note that, for all $K$'s in
Eqs.~(\ref{K1}) -- (\ref{K3}), $K_{\al\bbet}$ along the
configuration in \Eref{inftr} takes the form
\beq \label{Kab} \lf K_{\al\bbet}\rg=\diag\lf
M_\pm,\underbrace{K_{\gamma\bar\gamma},...,K_{\gamma\bar\gamma}}_{8~\mbox{\ftn
elements}}\rg~~~\mbox{with}~~~
M_\pm=\frac{1}{\fr^2}\mtta{\kappa}{\bar\kappa}{\bar\kappa}{\kappa}
~~~\mbox{and}~~~K_{\gamma\bar\gamma}=\begin{cases}
\fr^{-1}&\mbox{for}\>\>\>K=K_1\,,\\
1&\mbox{for}\>\>\>K=K_{2},~K_3\,.
\end{cases}\eeq
Here $\kp=\cm\fr^2-N\cp$ and $\bar\kp={N\cp^2\sg^2}$ where $N=3$
[$N=2$] for $K=K_1$ [$K=K_2$ or $K_3$]. Upon diagonalization of
$M_\pm$ we find its eigenvalues which are
\beq
\label{kpm}\kp_+=\cm\lf1+N\rpm(\cp\sg^2-1)/{\fr^2}\rg\simeq\cm
\>\>\>\mbox{and}\>\>\> \kp_-=\cm\lf1- {N\rpm}/{\fr}\rg,\eeq
where the positivity of $\kp_-$ is assured during and after HI for
\beq \label{rsmin}\rs<\fr/N~~\mbox{with}~~\rs=\cp/\cm\,.\eeq
Given that $\fr>1$ and $\vev{\fr}\simeq1$, \Eref{rsmin} implies
that the maximal possible $\rs$ is $\rs^{\rm max}\simeq1/N$. The
inequality above discriminates somehow the allowed parameter space
for the various choices of $K$'s in Eqs.~(\ref{K1}) -- (\ref{K2}).

Inserting \eqs{hpar}{Kab} into the kinetic term of the SUGRA
action, $K_{\al\bbet}\dot z^\al\dot z^\bbet$, we can specify the
canonically normalized fields, denoted by hat, as follows
\beq \label{VJe} \frac{d\se}{d\sg}=J,~~\widehat{\theta}_+
={J\over\sqrt{2}}\sg\theta_+,~~\widehat{\theta}_-
=\sqrt{\frac{\kp_-}{2}}\sg\theta_-,~~\widehat \theta_\Phi =
\sqrt{\kp_-}\sg\lf\theta_\Phi-\frac{\pi}{4}\rg~~~\mbox{and}~~~(\what{x}^{\gamma},\what{\bar
x}^{\gamma})=\sqrt{K_{\gamma\bar\gamma}}(x^\gamma,\bar
x^\gamma)\,,\eeq 
where $J=\sqrt{\kp_+}$ and
$\th_{\pm}=\lf\bar\th\pm\th\rg/\sqrt{2}$. As we show below, the
masses of the scalars besides $\se$ during HI are heavy enough
such that the dependence of the hatted fields on $\sg$ does not
influence their dynamics.

\subsection{Stability and one-Loop Radiative Corrections}\label{fhi11}

\renewcommand{\arraystretch}{1.2}
\begin{table}[!t]
\bec\begin{tabular}{|c|c|c|c|c|}\hline
{\sc Eigen-} & \multicolumn{4}{c|}{\sc Masses
Squared}\\\cline{2-5}
{\sc states}&& {$K=K_1$}&{$K=K_2$}&{$K=K_{3}$} \\
\hline\hline
$\widehat\theta_{+}$&$\widehat m_{\theta+}^2$&
\multicolumn{2}{|c|}{$6\Hhi^2$} &$6(1+1/\nsu)\Hhi^2$\\\cline{3-5}
$\widehat \theta_\Phi$ &$\widehat m_{ \theta_\Phi}^2$&
\multicolumn{3}{|c|}{$M^2_{BL}+\widehat
m_{\theta+}^2$}\\\cline{3-5}
$\widehat s, \widehat{\bar{s}}$ & $ \widehat m_{
s}^2$&$6\cp\sg^2\Hhi^2/N$&\multicolumn{2}{c|}{$6\Hhi^2/\nsu$}\\\cline{3-5}
$\widehat h_{\pm},\widehat{\bar h}_{\pm}$ &  $ \widehat
m_{h\pm}^2$&$3\Hhi^2\lf1\pm{4\lm}(\sg^{-2}+\cp)/{\ld}\rg+\widehat
m_{
s}^2/2$&\multicolumn{2}{c|}{$3\Hhi^2\lf1+1/\nsu\pm{4\lm}/{\ld\sg^2}\rg$}\\\cline{3-5}
$\widehat{\tilde\nu^c_{i}}, \widehat{\bar{\tilde\nu}}^c_{i}$ &  $
\widehat m_{i\tilde
\nu^c}^2$&$3\Hhi^2\lf1+{16\ld^2_{iN^c}}(\sg^{-2}+\cp)/{\ld^2}\rg+\widehat
m_{ s}^2/2$&\multicolumn{2}{c|}{$3\Hhi^2\lf1+1/\nsu+16\ld^2_{iN^c
}/\ld^2\sg^2\rg$}\\\cline{2-5}
$A_{BL}$ &  $ M_{BL}^2$&\multicolumn{3}{c|}{$g^2\cm\lf1-N\rpm
/\fr\rg\sg^2$}\\\hline
$\what \psi_\pm $ & $\what m^2_{ \psi\pm}$ &
$6\lf(N-3)\cp\sg^2-2\rg^2\Hhi^2/\cm\sg^2\fr^{2}$&\multicolumn{2}{c|}{$6\lf(N-2)\cp\sg^2-2\rg^2\Hhi^2/\cm\sg^2\fr^{2}$}\\
\cline{3-5} ${\widehat N_i^c}$& $\widehat
m_{{iN^c}}^2$&\multicolumn{3}{c|}{$48\ld^2_{iN^c
}\Hhi^2/\ld^2\sg^2$}\\\cline{2-5} $\ldu_{BL},
\widehat\psi_{\Phi-}$&
$M_{BL}^2$&\multicolumn{3}{c|}{$g^2\cm\lf1-N\rpm /\fr\rg\sg^2$}\\
\hline
\end{tabular}\eec
\hfill \caption{\sl\small The mass squared spectrum of our models
along the path in Eq.~(3.3) for $M\ll\sg\ll1$ and $N=3$ $[N=2]$
for $K=K_1$ [$K=K_2$ and $K_3$]. }\label{tab3}
\end{table}
\renewcommand{\arraystretch}{1.}

We can verify that the inflationary direction in \Eref{inftr} is
stable w.r.t the fluctuations of the non-inflaton fields. To this
end, we construct the mass-squared spectrum of the scalars taking
into account the canonical normalization of the various fields in
\Eref{VJe}. In the limit $\cm\gg\cp$, we find the expressions of
the masses squared $\what m^2_{z^\al}$ (with
$z^\al=\theta_+,\theta_\Phi,x^\gamma$ and $\bar x^\gamma$)
arranged in \Tref{tab3}. These results approach rather well for
$\sg=\sgx$ -- see \Sref{fhi2} -- the quite lengthy, exact
expressions taken into account in our numerical computation. The
various unspecified there eigenvalues are defined as follows
\beqs\beq \widehat h_\pm=(\widehat h_u\pm{\widehat
h_d})/\sqrt{2},~~~ \widehat{\bar h}_\pm=(\widehat{\bar
h}_u\pm\widehat{\bar h}_d)/\sqrt{2}~~~\mbox{and}~~~\what \psi_\pm
=(\what{\psi}_{\Phi+}\pm \what{\psi}_{S})/\sqrt{2}, \eeq
where the (unhatted) spinors $\psi_\Phi$ and $\psi_{\bar\Phi}$
associated with the superfields $\Phi$ and $\bar\Phi$ are related
to the normalized (hatted) ones in \Tref{tab3} as follows
\beq \label{psis}
\what\psi_{\Phi\pm}=\sqrt{\kp_\pm}\psi_{\Phi\pm}~~~
\mbox{with}~~~\psi_{\Phi\pm}=(\psi_\Phi\pm\psi_{\bar\Phi})/\sqrt{2}\,.
\eeq\eeqs

From \Tref{tab3} it is evident that $0<\nsu\leq6$ assists us to
achieve $m^2_{{s}}>\Hhi^2=\Vhi/3$ -- in accordance with the
results of \cref{su11} -- and also enhances the ratios
$m^2_{X^{\tilde\gamma}}/\Hhi^2$ for
$X^{\tilde\gamma}=\hu,\hd,\ssni$ w.r.t the values that we would
have obtained, if we had used just canonical terms in the $K$'s.
On the other hand, $\what m^2_{h-}>0$ requires
\beqs\bea \label{lm1} &\lm<\ld(1+\cp\sg^2/2)/4\lf1/\sg^2+\cp\rg
&~~\mbox{for}~~K=K_1;\\
\label{lm2} & \lm<\ld\sg^2(1+1/\nsu)/4&
~~\mbox{for}~~K=K_2~~\mbox{and}~~K_3\,.\eea\eeqs
In both cases, the quantity in the r.h.s of the inequality takes
its minimal value at $\sg=\sgf$ -- see \Sref{fhi2} -- and
numerically equals to $2\cdot10^{-5}-10^{-6}$. In \Tref{tab3} we
display also the mass $M_{BL}$ of the gauge boson which becomes
massive having `eaten' the Goldstone boson $\th_-$. This signals
the fact that $\Ggut$ is broken during HI and so no cosmological
defects are produced. Also, we can verify \cite{univ} that
radiative corrections \'a la Coleman-Weinberg can be kept under
control.

\subsection{Inflationary Observables}\label{fhi2}

A period of slow-roll HI is controlled by the strength of the
slow-roll parameters
\beq\label{sr}\widehat\epsilon= {1\over2}\left(\frac{\Ve_{{\rm
HI},\se}}{\Vhi}\right)^2\simeq \frac{8}{\cm\sg^2
\fr^{2}}~~~~\mbox{and}~~~~\widehat\eta = \frac{\Ve_{{\rm
HI},\se\se}}{\Vhi} \simeq12\:\frac{1 - \cp\sg^2}{\cm\sg^2
\fr^{2}}\,\cdot\eeq
Expanding $\widehat\epsilon$ and $\widehat\eta$ for $\sg\ll 1$ we
can find that HI terminates for $\sg=\sgf$ such that
\beq{\ftn\sf max}\{\widehat\epsilon(\sgf),|\widehat\eta(\sgf)|\}=1
~~~\Rightarrow~~~\sgf\simeq\mbox{\ftn\sf max}\left\{\frac{2
\sqrt{2/\cm}}{\sqrt{1+16\rs}},\frac{2
\sqrt{3/\cm}}{\sqrt{1+36\rs}}\right\}. \label{sgap}\eeq

The number of e-foldings, $\Ns$, that the pivot scale
$\ks=0.05/{\rm Mpc}$ suffers during HI can be calculated through
the relation
\begin{equation}
\label{Nhi}  \Ns=\:\int_{\se_{\rm f}}^{\se_\star}\, d\se\:
\frac{\Ve_{\rm HI}}{\Ve_{\rm HI,\se}}\simeq
\frac1{16\rs}\lf{(1+\cp\sgx^2)^{2}-1}\rg
\end{equation}
where $\sex$ [$\sgx$] is the value of $\se$ [$\sg$] when $\ks$
crosses the inflationary horizon. Given that $\sgf\ll\sgx$, we can
write $\sgx$ as a function of $\Ns$ as follows
\begin{equation}
\label{sgxb}\sgx\simeq {\sqrt{(\frs-1)/\cp}}~~\mbox{with}~~\frs=
\lf1+16\rs\Ns\rg^{1/2}
\end{equation}
We can impose a lower bound on $\cm$ above which $\sgx\leq1$ for
every $\rs$. Indeed, from \Eref{sgxb} we have
\begin{equation}
\label{cmmin}\sgx\leq1~~\Rightarrow~~\cm\geq{\lf\fns-1\rg}/{\rs}
\end{equation}
and so, our proposal can be stabilized against corrections from
higher order terms of the form $(\phc\phcb)^p$ with $p>1$ in
$\Whi$ -- see \Eref{Whi}. Despite the fact that $\cm$ may take
relatively large values, the corresponding effective theory is
valid up to $\mP=1$. To clarify further this point we have to
identify the ultraviolet cut-off scale $\Qef$ of theory analyzing
the small-field behavior of our models.  More specifically, we
expand about $\vev{\phi}=M\ll1$ the kinetic term $J^2 \dot\phi^2$
in the SUGRA action \cite{univ} and $\Vhi$ in \Eref{Vhi}. Our
results can be written in terms of $\se$ as
\beq\label{JVexp}  J^2
\dot\phi^2\simeq\lf1+3N\rs^2\what{\sg}^2-5N\rs^3\what{\sg}^4+\cdots\rg\dot\se^2~~~\mbox{and}~~~
\Vhi\simeq\frac{\ld^2\what{\sg}^4}{16\cm^{2}}\lf1-2\rs\what{\sg}^{2}+3\rs^2\what{\sg}^4-\cdots\rg\,.\eeq
From the expressions above we conclude that $\Qef=\mP$ since
$\rs\leq1$ due to \Eref{rsmin}.

The power spectrum $\As$ of the curvature perturbations generated
by $\sg$ at the pivot scale $\ks$ is estimated as follows
\beq \label{Proba}\sqrt{\As}=\: \frac{1}{2\sqrt{3}\, \pi} \;
\frac{\Ve_{\rm HI}(\sex)^{3/2}}{|\Ve_{\rm
HI,\se}(\sex)|}\simeq\frac{
\ld\sqrt{\cm}}{32\sqrt{3}\pi}\sgx^3~~~\Rightarrow~~~\ld=32\sqrt{3\As}\pi
\cm\lf\frac{\rs}{\frs-1}\rg^{3/2}\,\cdot  \eeq
The resulting relation reveals that $\ld$ is proportional to $\cm$
for fixed $\rs$.

At the same pivot scale, we can also calculate $\ns$, its running,
$\as$, and $r$ via the relations
\beqs\baq \label{ns} && \ns=\: 1-6\widehat\epsilon_\star\ +\
2\widehat\eta_\star\simeq
1-\frac{3}{2\Ns}-\frac{3}{8(\Ns^3\rs)^{1/2}}\,,~~r=16\widehat\epsilon_\star\simeq
\frac{1}{2\Ns^2\rs}+\frac{2}{(\Ns^3\rs)^{1/2}}\,,\\
&& \label{as} \as =\:{2\over3}\left(4\widehat\eta_\star^2-(n_{\rm
s}-1)^2\right)-2\widehat\xi_\star\simeq-\frac{3}{2\Ns^2}
~~\mbox{with}~~ \widehat\xi={\Ve_{\rm HI,\widehat\sg} \Ve_{\rm
HI,\widehat\sg\widehat\sg\widehat\sg}/\Ve_{\rm HI}^2}\,.\eaq\eeqs
Here the variables with subscript $\star$ are evaluated at
$\sg=\sgx$. A clear dependence of $\ns$ and $r$ on $\rs$ arises.


\subsection{Comparison with Observations}\label{fhi3}

The approximate analytic expressions above can be verified by the
numerical analysis of our model. Namely, we apply the accurate
expressions in \eqs{Nhi}{Proba} and confront the corresponding
observables with the requirements \cite{plin,plcp}
\begin{equation}
\label{Ntot} \mbox{\ftn\sf
(a)}\>\>\Ns\simeq61.5+\ln{\Vhi(\sgx)^{1/2}\over\Vhi(\sgf)^{1/4}}+\frac12\fr(\sgx)~~~\mbox{and}~~~\mbox{\ftn\sf
(b)}\>\>\As^{1/2}\simeq4.627\cdot10^{-5}\,,\eeq
where we consider in \sEref{Ntot}{a} an equation-of-state
parameter $w_{\rm int}=1/3$ correspoding to quartic potential
which is expected to approximate rather well $\Vhi$ for $\sg\ll1$.
We, thus, restrict $\ld$ and $\sgx$ and compute the model
predictions via \eqs{ns}{as} for any selected $\rs$. These must be
in agreement with the fitting of the \plk, \emph{Baryon Acoustic
Oscillations} ({\sf\ftn BAO}) and \bcp\ data \cite{plin,gwsnew}
with $\Lambda$CDM$+r$ model, i.e.,
\begin{equation}  \label{nswmap}
\mbox{\ftn\sf
(a)}\>\>\ns=0.968\pm0.009\>\>\>~\mbox{and}\>\>\>~\mbox{\ftn\sf
(b)}\>\>r\leq0.07,
\end{equation}
at 95$\%$ \emph{confidence level} ({\sf\ftn c.l.}) with
$|\as|\ll0.01$.

\begin{figure}[!t]\vspace*{-.12in}
\hspace*{-.19in}
\begin{minipage}{8in}
\epsfig{file=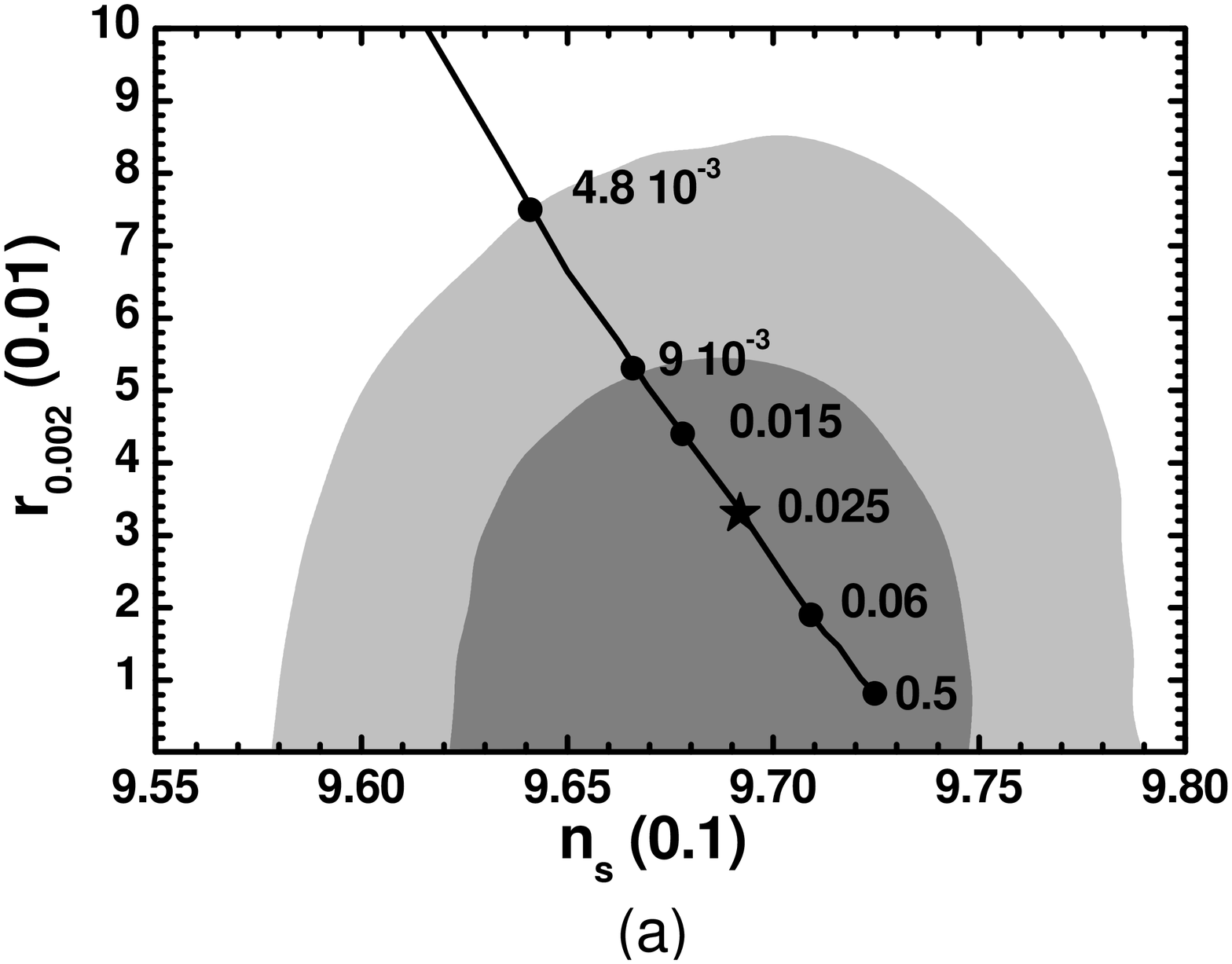,height=3.6in,angle=-90}
\hspace*{-1.2cm} \epsfig{file=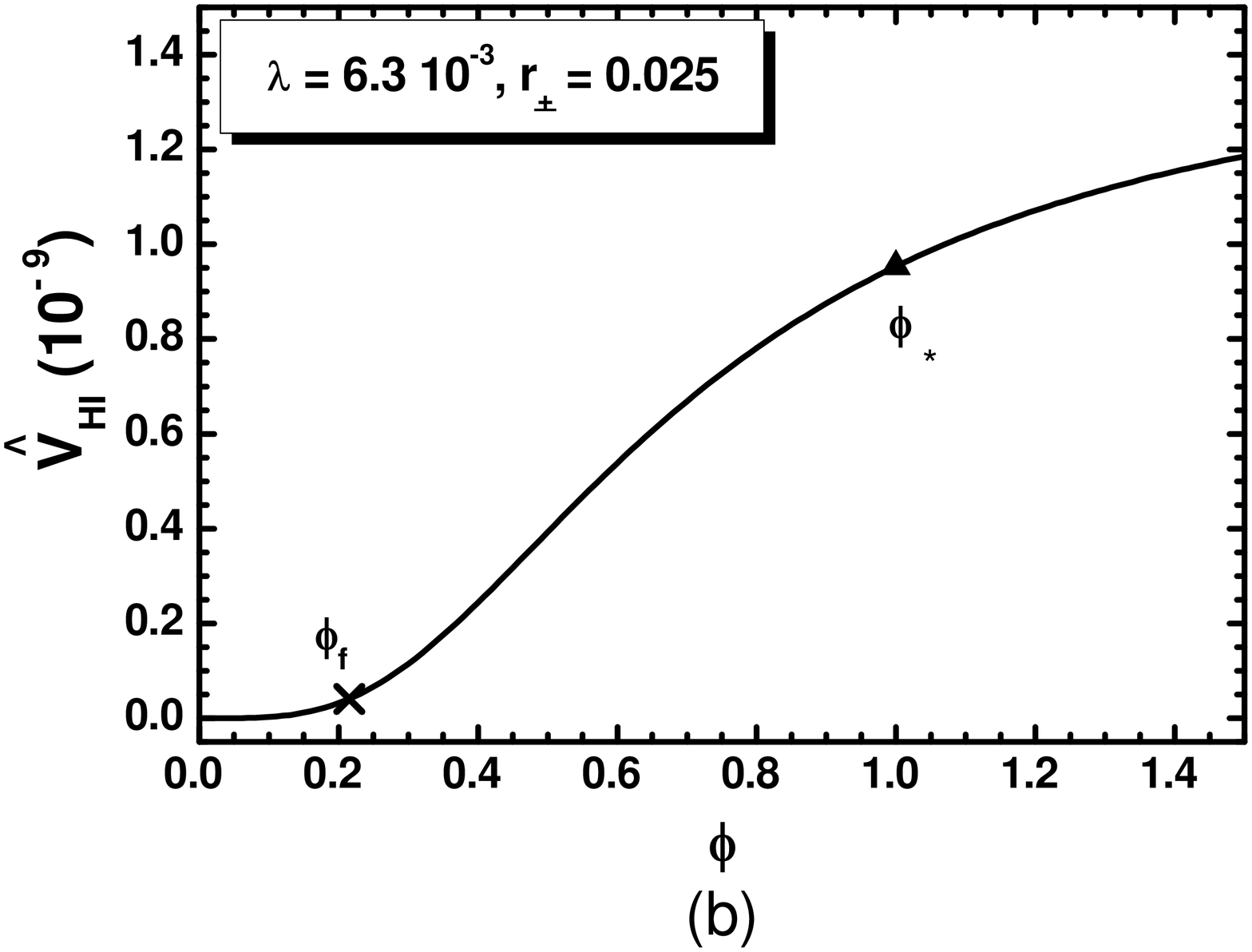,height=3.6in,angle=-90}
\hfill
\end{minipage}
\hfill \caption{\sl\small {\sffamily\ftn (a)} Allowed curve in the
$\ns-\rw$ plane for $K=K_2$ or $K_3$ with the $\rs$ values
indicated on it -- the marginalized joint $68\%$ [$95\%$] regions
from \plk, BAO and BK14 data \cite{gwsnew} are depicted by the
dark [light] shaded contours; {\sffamily\ftn (b)} The inflationary
potential $\Vhi$ as a function of $\sg$ for $\sg>0$,
$\rs\simeq0.025$ and $\ld=6.3\cdot 10^{-3}$ -- the values of
$\sgx$, $\sgf$ and are also indicated.}\label{fig1}
\end{figure}\renewcommand{\arraystretch}{1.}


Let us clarify here that the free parameters of our models are
$\rs$ and $\ld/\cm$ and not $\cm$, $\cp$ and $\ld$ as naively
expected. Indeed, if we perform the rescalings
\beq\label{resc}
\Phi\to\Phi/\sqrt{\cm},~~\bar\Phi\to\bar\Phi/\sqrt{\cm}~~~\mbox{and}~~~
S\to S,\eeq
$W$ in \Eref{Whi} depends on $\ld/\cm$ and the $K$'s in \Eref{K1}
-- (\ref{K3}) depend on $\rs$. As a consequence, $\Vhi$ depends
exclusively on $\ld/\cm$ and $\rs$. Since the $\ld/\cm$ variation
is rather trivial -- see \cref{nMHkin} -- we focus on the
variation of the other parameters.

Our results are displayed in \Fref{fig1} for $K=K_2$ or $K_3$.
Namely, in \sFref{fig1}{a} we show a comparison of the models'
predictions against the observational data \cite{plin,gwsnew} in
the $\ns-\rw$ plane, where $\rw=16\eph(\se_{0.002})$ with
$\se_{0.002}$ being the value of $\se$ when the scale
$k=0.002/{\rm Mpc}$, which undergoes $\what N_{0.002}=\Ns+3.22$
e-foldings during HI, crosses the horizon of HI. We depict the
theoretically allowed values with a solid line with the variation
of $\rs$ shown along it. For low enough $\rs$'s -- i.e.
$\rs\leq0.0005$ -- the line reaches $(\ns,\rw)\simeq(0.947,0.28)$
obtained within \emph{minimal} quartic inflation defined for
$\cp=0$. Increasing $\rs$ the line enters the observationally
allowed regions and terminates for $\rs\simeq0.5$, beyond which
\Eref{rsmin} is violated. Along this line we find -- consistently
with the analytic formulas of \Sref{fhi2}
\beq\label{res1} 0.048\lesssim {\rs\over0.1}\lesssim5,\>\>\>
9.64\lesssim {\ns\over0.1}\lesssim9.72,\>\>\> 0.7\lesssim
{r\over0.01}\lesssim8.1\>\>\>\mbox{and}\>\>\>0.17\lesssim
10^5{\ld\over \cm}\lesssim3.13\,. \eeq  Moreover
$\as\simeq-(5-6)\cdot10^{-4}$ and so, our models are consistent
with the fitting of data with the $\Lambda$CDM+$r$ model
\cite{plin}. These are also testable by the forthcoming
experiments, like {\sc Bicep3}, PRISM and LiteBIRD, searching for
primordial gravity waves since $r\gtrsim0.007$. Had we employed
$K=K_1$, the line in \sFref{fig1}{a} would have been shortened
until $\rs\simeq0.33$ yielding $\rw\gtrsim0.0084$. The other
bounds would have been remained more or less unaffected.

Taking the $\chi^2$ distribution of the obtained $(\ns,r)$'s we
can identify the following best-fit value:
\beq \rs=0.025~~\mbox{resulting
to}~~(\ns,r)=(0.969,0.033)\,.\label{res2} \eeq For this value we
display the structure of $\Vhi$ as a function of $\sg$ in
\sFref{fig1}{b}. We take $\sgx=1$ which corresponds to
$\ld=6.3\cdot 10^{-3}$ and $\cm=146$. We observe that $\Vhi$ is a
monotonically increasing function of $\sg$. The inflationary
scale, $\Vhi^{1/4}$, approaches the SUSY GUT scale $M_{\rm
GUT}\simeq 8.2\cdot10^{-3}$ and lies well below $\Qef=1$,
consistently with the classical approximation to the inflationary
dynamics.

\section{Higgs Inflation and $\mu$ Term of MSSM}
\label{secmu}

A byproduct of the $R$ symmetry associated with our model is that
it assists us to understand the origin of $\mu$ term of MSSM, as
we show in \Sref{secmu1}, consistently with the low-energy
phenomenology of MSSM  -- see \Sref{pheno}. Hereafter we restore
units, i.e., we take $\mP=2.433\cdot10^{18}~\GeV$.

\subsection{SUSY Potential}\label{secmu1}

The SUSY limit $V_{\rm SUSY}$ of $\Vhi$ in \Eref{Vhi} is given by
\beqs \beq \label{Vsusy} V_{\rm SUSY}= \widetilde K^{\al\bbet}
W_{\rm HI\al} W^*_{\rm HI\bbet}+\frac{g^2}2 \mbox{$\sum_{\rm a}$}
{\rm D}_{\rm a} {\rm D}_{\rm a}\,,\eeq
where $\widetilde K$ is the limit of the $K$'s in Eqs.~(\ref{K1})
-- (\ref{K3}) for $\mP\to\infty$. Focusing on the $S-\phcb-\phc$
system we find
\beq \label{Kquad}\widetilde K=\cm F_- -N\cp F_+ +|S|^2.\eeq
Upon substitution of $\widetilde K$ into \Eref{Vsusy} we obtain
\bea \nonumber && V_{\rm
SUSY}=\ld^2\left|\phcb\phc-\frac14{M^2}\right|^2 +\frac{\ld^2
}{\cm(1-N\rs)}S^2\lf|\phc|^2+|\phcb|^2\rg+\frac{g^2}2\cm^2(1-N\rs)^2\lf|\phc|^2-|\phcb|^2\rg^2\,.
\label{VF}\eea\eeqs  From the last equation, we find that the SUSY
vacuum lies along the D-flat direction $|\phcb|=|\phc|$ with
\beq \vev{S}=0 \>\>\>\mbox{and}\>\>\>
|\vev{\Phi}|=|\vev{\bar\Phi}|=M/2\,.\label{vevs} \eeq
As a consequence, $\vev{\Phi}$ and $\vev{\bar\Phi}$ break
spontaneously $U(1)_{B-L}$ down to $\mathbb{Z}^{B-L}_2$. Since
$U(1)_{B-L}$ is already broken during HI, no cosmic string are
formed.

\subsection{Generation of the $\mu$ Term of MSSM}\label{secmu2}

The contributions from the soft SUSY breaking terms, although
negligible during HI, since these are much smaller than $\sg$, may
shift slightly $\vev{S}$ from zero in \Eref{vevs}. Indeed, the
relevant potential terms are
\beq V_{\rm soft}= \lf\ld A_\ld S \phcb\phc+\lm A_\mu S \hu\hd +
\ld_{iN^c} A_{iN^c}\phc \widetilde N^{c2}_i- {\rm a}_{S}S\ld M^2/4
+ {\rm h. c.}\rg+ m_{\gamma}^2\left|X^\gamma\right|^2,
\label{Vsoft} \eeq
where $m_{\gamma}, A_\ld, A_\mu, A_{iN^c}$ and $\aS$ are soft SUSY
breaking mass parameters.  Rotating $S$ in the real axis by an
appropriate $R$-transformation, choosing conveniently the phases
of $\Ald$ and $\aS$ so as the total low energy potential $V_{\rm
tot}=V_{\rm SUSY}+V_{\rm soft}$ to be minimized -- see \Eref{VF}
-- and substituting in $V_{\rm soft}$ the SUSY \emph{vacuum
expectation values} ({\sf\ftn v.e.vs}) of $\phc$  and $\phcb$ from
\Eref{vevs} we get
\beqs\beq \vev{V_{\rm tot}(S)}= \ld^2\,M^2S^2/2\cm(1-N\rs)-\ld\am
\mgr M^2 S, \label{Vol} \eeq
where we take into account that $m_S\ll M$ and we set $|A_\ld| +
|{\rm a}_{S}|=2\am\mgr$ with $\mgr$ being the $\Gr$ mass and
$\am>0$ a parameter of order unity which parameterizes our
ignorance for the dependence of $|A_\ld|$ and $|{\rm a}_{S}|$ on
$\mgr$.  The minimization condition for the total potential in
\Eref{Vol} w.r.t $S$ leads to a non vanishing $\vev{S}$ as follows
\beq \label{vevS}\frac{d}{d S} \vev{V_{\rm
tot}(S)}=0~~\Rightarrow~~\vev{S}\simeq
\am\mgr\cm(1-N\rs)/\ld.\eeq\eeqs
The generated $\mu$ term from the second term in the r.h.s of
\Eref{Whi} is \beq\mu =\lm \vev{S}
\simeq\lm\am\mgr\cm(1-N\rs)/\ld\,.\label{mu}\eeq
By virtue of \Eref{Proba}, the resulting $\mu$ above depends on
$\rs$ and does not depend on $\ld$ and $\cm$. We may verify that
any $|\mu|$ value is accessible for the $\lm$ values allowed by
\eqs{lm1}{lm2} without any ugly hierarchy between $\mgr$ and
$\mu$.

\subsection{Connection with the MSSM Phenomenology}\label{pheno}

The SUSY breaking effects, considered in \Eref{Vsoft}, explicitly
break $U(1)_R$ to a subgroup, $\mathbb{Z}_2^{R}$ which can be
identified with a matter parity. Under this discrete symmetry all
the matter (quark and lepton) superfields change sign -- see
Table~1. From the $R$ charges there we conclude that
$\mathbb{Z}_2^{R}$ remains unbroken, since $\vev{S}$ in
\Eref{vevS} also breaks spontaneously $U(1)_R$ to
$\mathbb{Z}_2^{R}$ and so no disastrous domain walls are formed.
Combining $\mathbb{Z}_2^{R}$ with the $\mathbb{Z}_2^{\rm f}$
fermion parity, under which all fermions change sign, yields the
well-known $R$-parity. This residual symmetry prevents rapid
proton decay and guarantees the stability of the \emph{lightest
SUSY particle} ({\sf\ftn LSP}), providing thereby a well-motivated
\emph{cold dark matter} ({\sf\ftn CDM}) candidate.

The candidacy of LSP may be successful, if it generates the
correct CDM abundance \cite{plcp} within a concrete low energy
framework. In our case this is the MSSM or, more specifically, the
\emph{Constrained MSSM} ({\ftn\sf CMSSM}), if we adopt only the
following free parameters
\begin{equation}
{\rm
sign}\mu,~~\tan\beta=\vev{\hu}/\vev{\hd},~~\mg,~~m_0,~~\mbox{and}~~A_0,
\label{para}
\end{equation}
where ${\rm sign}\mu$ is the sign of $\mu$, and the three last
mass parameters denote the common gaugino mass, scalar mass and
trilinear coupling constant, respectively, defined (normally) at
$\Mgut$. The parameter $|\mu|$ is not free, since it is computed
at low scale by enforcing the conditions for the electroweak
symmetry breaking. The values of the (four and one half)
parameters in \Eref{para} can be tightly restricted imposing a
number of cosmo-phenomenological constraints from which the
consistency of LSP relic density with observations plays a central
role. Some updated results are recently presented in \cref{mssm},
where we can also find the best-fit values of $|A_0|$, $m_0$ and
$|\mu|$ listed in the first four lines of \Tref{tab}.  We see that
there are four allowed regions characterized by the specific
mechanism for suppressing the relic density of the LSP which is
the lightest neutralino ($\chi$) -- $\tilde\tau_1, \tilde t_1$ and
$\tilde \chi^\pm_1$ stand for the lightest stau, stop and chargino
eigenstate. If we take the best-fit value of $\rs$ in \Eref{res2}
and identify
\beq m_0=\mgr~~\mbox{and}~~|A_0|=|A_\ld|=|\aS|\label{softass}\eeq
we can derive first $\am$ and then the $\lm$ values which yield
the phenomenologically desired $|\mu|$ -- ignoring renormalization
group effects. The outputs of our computation is listed in the two
rightmost columns of \Tref{tab} for $K=K_1, K_2$ and $K_3$. From
these we infer that the required $\lm$ values, in all cases
besides the one, written in italics, are comfortably compatible
with \eqs{lm1}{lm2} for $\nsu=2$ which imply
\beq
\lm\lesssim6.6\cdot10^{-6}~~\mbox{for}~~K=K_1~~\mbox{and}~~\lm\lesssim1.1\cdot10^{-5}~~\mbox{for}~~
K=K_2~~\mbox{and}~~K_3\,.\label{lmumax}\eeq
Therefore, we conclude that the whole inflationary scenario can be
successfully combined with all the allowed regions CMSSM besides
the $\tilde\tau_1-\chi$ coannihilation region for $K=K_1$. On the
other hand, all the CMSSM regions can be consistent with the
gravitino limit on $\Trh$ -- see \Sref{lept1}. Indeed, $\mgr$ as
low as $1~\TeV$ becomes cosmologically safe, under the assumption
of the unstable $\Gr$, for the $\Trh$ values, necessitated for
satisfactory leptogenesis, as presented in \Tref{tab2}.

\renewcommand{\arraystretch}{1.25}
\begin{table}[!t] \bec
\begin{tabular}{|c|c|c|c||c|c|c|}\hline
{\sc CMSSM}&$|A_0|$&$m_0$&$|\mu| $&$\am$&\multicolumn{2}{|c|}{\sc
$\lm~(10^{-6})$}\\\cline{6-7} {\sc Region
}&$(\TeV)$&$(\TeV)$&$(\TeV)$&&$K=K_1$&$K=K_2,~K_3$\\\hline\hline
$A/H$ Funnel &$9.9244$ &$9.136$&$1.409$&$1.086$ &$0.6223$&$0.607$\\
$\tilde\tau_1-\chi$ Coannihilation &$1.2271$ &$1.476$&$2.62$& $0.831$&{\it 9.36}&$9.12$\\
$\tilde t_1-\chi$ Coannihilation  &$9.965$ &$4.269$&$4.073$&$2.33$ &$1.794$&$1.75$\\
$\tilde \chi^\pm_1-\chi$ Coannihilation  &$9.2061$ &$9.000$&$0.983$&$1.023$ &$0.468$&$0.456$\\
\hline
\end{tabular}
\end{center}
\caption[]{\sl\small The required $\lm$ values which render our
models compatible with the best-fit points in the CMSSM, as found
in \cref{mssm}, for the assumptions of \Eref{softass} $K=K_1$ or
$K=K_2$ and $K_3$ with $\nsu=2$ and $\rs=0.025$.} \label{tab}
\end{table}\renewcommand{\arraystretch}{1.}

\section{Non-Thermal Leptogenesis and Neutrino Masses}\label{pfhi}

We below specify how our inflationary scenario makes a transition
to the radiation dominated era (\Sref{lept0}) and offers an
explanation of the observed BAU (\Sref{lept1}) consistently with
the $\Gr$ constraint and the low energy neutrino data. Our results
are summarized in \Sref{num}.

\subsection{Inflaton Mass \& Decay}\label{lept0}

When HI is over, the inflaton continues to roll down towards the
SUSY vacuum, \Eref{vevs}. Soon after, it settles into a phase of
damped oscillations around the minimum of $\Vhi$. The (canonically
normalized) inflaton,
\beq\dphi=\vev{J}\dph\>\>\>\mbox{with}\>\>\> \dph=\phi-M
\>\>\>\mbox{and}\>\>\>\vev{J}=\sqrt{\vev{\kp_+}}=\sqrt{\cm(1-{N\rs})}\label{dphi}
\eeq
acquires mass, at the SUSY vacuum in \Eref{vevs}, which is given
by
\beq \label{msn} \msn=\left\langle\Ve_{\rm
HI,\se\se}\right\rangle^{1/2}= \left\langle \Ve_{\rm
HI,\sg\sg}/J^2\right\rangle^{1/2}\simeq\frac{\ld
M}{\sqrt{2\cm\lf1-{N\rs}\rg}}\,,\eeq
where the last (approximate) equality above is valid only for
$\rs\ll1/N$ -- see \eqs{kpm}{VJe}.  As we see, $\msn$ depends
crucially on $M$ which may be, in principle, a free parameter
acquiring any subplanckian value without disturbing the
inflationary process.  To determine better our models, though, we
prefer to specify $M$ requiring that $\vev{\Phi}$ and
$\vev{\bar\Phi}$ in \Eref{vevs} take the values dictated by the
unification of the MSSM gauge coupling constants, despite the fact
that $U(1)_{B-L}$ gauge symmetry does not disturb this unification
and $M$ could be much lower. In particular, the unification scale
$\Mgut\simeq2\cdot10^{16}~\GeV$ can be identified with $M_{BL}$ --
see \Tref{tab3} -- at the SUSY vacuum in \Eref{vevs}, i.e.,
\beq \label{Mg} {\sqrt{\cm(\vev{\fr}-N\rs)}gM\over
\sqrt{\vev{\fr}}}=\Mgut\>\Rightarrow\>M\simeq{\Mgut}/{g\sqrt{\cm\lf1-{N\rs}\rg}}\eeq
with $g\simeq0.7$ being the value of the GUT gauge coupling and we
take into account that $\vev{\fr}\simeq1$. Upon substitution of
the last expression in \Eref{Mg} into \Eref{msn} we can infer that
$\msn$ remains constant for fixed $\rs$ since $\ld/\cm$ is fixed
too -- see \Eref{Proba}. Particularly, along the line in
\sFref{fig1}{a} we obtain
\beqs\bea \label{resmass} &3.5\cdot10^{11}\lesssim
{\msn/\GeV}\lesssim3.9\cdot10^{13}&~~\mbox{for}~~K=K_1;\\
&3.46\cdot10^{10}\lesssim
{\msn/\GeV}\lesssim4.2\cdot10^{13}&~~\mbox{for}~~K=K_2~~\mbox{and}~~K_3\,,
\eea\eeqs
%


During the phase of its oscillations at the SUSY vacuum, $\dphi$
decays perturbatively  reheating the Universe at a reheat
temperature given by
\beq\Trh=
\left({72/5\pi^2g_{*}}\right)^{1/4}\lf\Gsn\mP\rg^{1/2}\>\>\>\mbox{with}\>\>\>\Gsn=\GNsn+\Ghsn\,.\label{Trh}\eeq
Also $g_{*}=228.75$ counts the MSSM effective number of
relativistic degrees of freedom. To compute $\Trh$ we take into
account the following decay widths:
\beqs\bea \GNsn&=&\frac{g_{iN^c}^2}{16\pi}\msn\lf1-\frac{4\mrh[
i]^2}{\msn^2}\rg^{3/2}\>\>\mbox{with}\>\>
g_{iN^c}=\frac{\ld_{iN^c}}{\vev{J}}\lf1-3\cp
\frac{N}{2}\frac{M^2}{\mP^2}\rg\\
\Ghsn&=&\frac{2}{8\pi}g_{H}^2\msn\>\>\mbox{with}\>\>
g_{H}=\frac{\lm}{\sqrt{2}}\lf1-2\cp\frac{M^2}{\mP^2}\rg \eea
arising from the lagrangian terms
\bea {\cal L}_{\dphi\to \sni\sni}&=&-\frac12e^{K/2\mP^2}W_{{\rm
HI},N_i^cN^c_i}\sni\sni\ +\ {\rm h.c.}=g_{iN^c} \dphi\
\lf\sni\sni\ +\ {\rm h.c.}\rg +\cdots,\\ {\cal
L}_{\dphi\to\hu\hd}&=&-e^{K/\mP^2}K^{SS^*}\left|W_{{\rm
HI},S}\right|^2 =-g_{H} \msn\dphi \lf H_u^*H_d^*\ +\ {\rm
h.c.}\rg+\cdots\eea\eeqs
describing $\dphi$ decay into a pair of $N^c_{j}$ with Majorana
masses $\mrh[j]=\ld_{iN^c}M$ and $\hu$ and $\hd$ respectively.
Note that the decay modes into MSSM (s)-particles $XYZ$
\cite{univ} through a typical trilinear superpotential term of
MSSM is suppressed since they arise from non-renormalizable
interactions proportional to $M/\mP\ll1$.

\subsection{Lepton-Number and Gravitino Abundances}\label{lept1}

For $\Trh<\mrh[i]$, the out-of-equilibrium decay of $N^c_{i}$
generates a lepton-number asymmetry (per $N^c_{i}$ decay),
$\ve_i$. The resulting lepton-number asymmetry is partially
converted through sphaleron effects into a yield of the observed
BAU:
\beq Y_B=-0.35\cdot{5\over2}{\Trh\over\msn}\mbox{$\sum_i$}
{\GNsn\over\Gsn}\ve_i,\>\>\>\mbox{with}\>\>\>\ve_i =\sum_{j\neq
i}\frac{
\im\left[(\mD[]^\dag\mD[])_{ij}^2\right]}{8\pi\vev{\hu}^2(\mD[]^\dag\mD[])_{ii}}\bigg(
F_{\rm S}\lf x_{ij},y_i,y_j\rg+F_{\rm
V}(x_{ij})\bigg),\label{Yb}\eeq
where $\vev{\hu}\simeq174~\GeV$, for large $\tan\beta$, $m_{\rm
D}$ the Dirac mass matrix of $\nu_i$ and $F_{\rm S}$ [$F_{\rm V}$]
are the functions entered in the vertex and self-energy
contributions computed as indicated in \cref{plum}. The expression
above has to reproduce the observational result \cite{plcp}:
\beq
Y_B=\lf8.64^{+0.15}_{-0.16}\rg\cdot10^{-11}.\label{BAUwmap}\eeq
The validity of \Eref{Yb} requires that the $\dphi$ decay into a
pair of $\sni$'s is kinematically allowed for at least one species
of the $\sni$'s and also that there is no erasure of the produced
$Y_L$ due to $N^c_1$ mediated inverse decays and $\Delta L=1$
scatterings. These prerequisites are ensured if we impose
\beq\label{kin} {\sf \ftn
(a)}\>\>\msn\geq2\mrh[1]\>\>\>\mbox{and}\>\>\>{\sf \ftn
(b)}\>\>\mrh[1]\gtrsim 10\Trh.\eeq
The quantity $\ve_i$ can be expressed in terms of the Dirac masses
of $\nu_i$, $\mD[i]$, arising from the third term of \Eref{Whi}.
Employing the seesaw formula we can then obtain the light-neutrino
mass matrix $m_\nu$ in terms of $\mD[i]$ and $\mrh[i]$. As a
consequence, nTL can be nicely linked to low energy neutrino data.
We take as inputs the recently updated best-fit values
\cite{valle} -- cf. \cref{univ} -- on the neutrino mass-squared
differences, $\Delta m^2_{21}=7.56\cdot10^{-5}~{\rm eV}^2$ and
$\Delta m^2_{31}=2.55\cdot10^{-3}~{\rm eV}^2$ $\left[\Delta
m^2_{31}=2.49\cdot10^{-3}~{\rm eV}^2\right]$, on the mixing
angles, $\sin^2\theta_{12}=0.321$,
$\sin^2\theta_{13}=0.02155~\left[\sin^2\theta_{13}=0.0214\right]$
and $\sin^2\theta_{23}=0.43~\left[\sin^2\theta_{23}=0.596\right]$
and the CP-violating Dirac phase
$\delta=1.4\pi~\left[\delta=1.44\pi\right]$ for \emph{normal
[inverted] ordered} (NO [IO]) \emph{neutrino masses}, $\mn[i]$'s.
Furthermore, the sum of $\mn[i]$'s is bounded from above at 95\%
c.l. by the data \cite{plcp}, \beq\mbox{$\sum_i$}
\mn[i]\leq0.23~{\eV}.\label{sumnu}\eeq

The required $\Trh$ in \Eref{Yb} must be compatible with
constraints on the gravitino ($\Gr$) abundance, $Y_{3/2}$, at the
onset of \emph{nucleosynthesis} (BBN), which is estimated to be:
\beq\label{Ygr} Y_{3/2}\simeq 1.9\cdot10^{-22}\ \Trh/\GeV ,\eeq
where we take into account only thermal production of $\Gr$, and
assume that $\Gr$ is much heavier than the MSSM gauginos. On the
other hand, $\Yg$  is bounded from above in order to avoid
spoiling the success of the BBN. For the typical case where $\Gr$
decays with a tiny hadronic branching ratio, we have
\beq  \label{Ygw} \Yg\lesssim\left\{\bem
%
10^{-14}\hfill \cr
10^{-13}\hfill \cr
10^{-12}\hfill \cr \eem
\right.\>\>\>\mbox{for}\>\>\>\mgr\simeq\left\{\bem
0.69~\TeV\hfill \cr
10.6~\TeV\hfill \cr
13.5~\TeV\hfill \cr \eem
\right.\>\>\>\mbox{implying}\>\>\>\Trh\lesssim5.3\cdot\left\{\bem
%
10^{7}~\GeV\,,\hfill \cr
10^{8}~\GeV\,,\hfill \cr
10^{9}~\GeV\,.\hfill \cr\eem
\right.\eeq
The bounds above can be somehow relaxed in the case of a stable
$\Gr$.

\subsection{Results}\label{num}

\renewcommand{\arraystretch}{1.3}
\begin{table}[!t]
\bec\begin{tabular}{|c||c|c||c|c|c||c|c|}\hline
{\sc Parameters} &  \multicolumn{7}{c|}{\sc Cases}\\\cline{2-8}
&A&B& C & D& E & F&G\\ \cline{2-8} &\multicolumn{2}{c||}{\sc
Normal} & \multicolumn{3}{c||}{\sc Almost}&
\multicolumn{2}{c|}{\sc Inverted}
\\& \multicolumn{2}{c||}{\sc Hierarchy}&\multicolumn{3}{c||}{\sc Degeneracy}&
\multicolumn{2}{c|}{\sc Hierarchy}\\ \hline  \hline
\multicolumn{8}{|c|}{\sc Low Scale Parameters}\\\hline
$\mn[1]/0.1~\eV$&$0.05$&$0.1$&$0.5$ & $0.7$& $0.7$ & $0.5$&$0.49$\\
$\mn[2]/0.1~\eV$&$0.1$&$0.13$&$0.51$ & $0.7$& $0.7$ & $0.51$&$0.5$\\
$\mn[3]/0.1~\eV$&$0.5$&$0.51$&$0.7$ & $0.86$&$0.5$ &
$0.1$&$0.05$\\\hline
$\sum_i\mn[i]/0.1~\eV$&$0.65$&$0.74$&$1.7$ & $2.3$&$1.9$ &
$1.1$&$1$\\ \hline
$\varphi_1$&$-\pi/5$&$-\pi/2$&$\pi$ & $\pi/9$&$0$ & $-3\pi/4$&$\pi/2$\\
$\varphi_2$&$\pi$&$0$ &$\pi/3$& $\pi$&$\pi/2$ &
$5\pi/4$&$-\pi/2$\\\hline
\multicolumn{8}{|c|}{\sc Leptogenesis-Scale Parameters}\\\hline
$\mD[1]/0.1~\GeV$&$2$&$5$&$10$ & $10$&$1.3$ & $5$&$6$\\
$\mD[2]/\GeV$&$6$&$1.97$&$3.9$ & $10$&$9$ & $0.715$&$1.1$\\
$\mD[3]/\GeV$&$100$&$150$&$170$ & $168$&$202$ &
$100$&$199$\\\hline
$\mrh[1]/10^{10}~\GeV$&$1.0$&$3.3$&$2.85$ & $3.3$&$2.98$ & $0.45$&$1.23$\\
$\mrh[2]/10^{10}~\GeV$&$6.9$&$13.6$&$26.5$ & $111.4$&$13.9$ & $2.76$&$2.8$\\
$\mrh[3]/10^{14}~\GeV$&$2.9$&$4.9$&$2.2$ &
$1.2$&$3.7$&$2.7$&$27.2$\\\hline
\multicolumn{8}{|c|}{\sc Open Decay Channels of the Inflaton,
\dphi, Into $\sni$}\\\hline
$\dphi\ \to$&$\wrhn[1]$&$\wrhn[1]$& $\wrhn[1]$& $\wrhn[1]$&
$\wrhn[1]$ & $\wrhn[1,2]$&$\wrhn[1,2]$\\ \hline
$\GNsn/\Gsn~(\%)$&$13.8$&$15.4$& $17.4$& $14.9$& $17.1$ & $18.3$&$22.7$\\
\hline
\multicolumn{8}{|c|}{\sc Resulting $B$-Yield }\\\hline
$10^{11}Y_B$&$8.68$&$8.66$&$8.79$ & $8.69$&$8.58$ &
$8.67$&$8.68$\\\hline
\multicolumn{8}{|c|}{\sc Resulting $\Trh$ and $\Gr$-Yield
}\\\hline
$\Trh/10^{7}~\GeV$&$2.8$&$2.8$&$2.84$ & $2.8$&$2.84$ & $2.85$&$2.94$\\
$10^{15}\Yg$&$5.3$&$5.3$&$5.4$ & $5.3$&$5.4$ & $5.4$&$5.5$\\\hline
\end{tabular}\eec
\hfill \caption[]{\sl\small  Parameters yielding the correct $Y_B$
for various neutrino mass schemes. We take $K=K_2$ or $K_3$ with
$\nsu=2$, $\rs=0.025$ and $\lm=10^{-6}$.} \label{tab2}
\end{table}

Confronting with observations $Y_B$ and $\Yg$ which depend on
$\msn$, $\Trh$, $\mrh[i]$ and $\mD[i]$'s  -- see \eqs{Yb}{Ygr} --
we can further constrain the parameter space of the our models. We
follow the bottom-up approach detailed in \cref{univ}, according
to which we find the $\mrh[i]$'s by using as inputs the
$\mD[i]$'s, a reference mass of the $\nu_i$'s -- $\mn[1]$ for NO
$\mn[i]$'s, or $\mn[3]$ for IO $\mn[i]$'s --, the two Majorana
phases $\varphi_1$ and $\varphi_2$ of the PMNS matrix, and the
best-fit values for the low energy parameters of neutrino physics
mentioned in \Sref{lept1}. In our numerical code, we also estimate
\cite{univ} the RG evolved values of the latter parameters at the
scale of nTL, $\Lambda_L=\msn$, by considering the MSSM with
$\tan\beta\simeq50$ as an effective theory between $\Lambda_L$ and
the soft SUSY breaking scale, $M_{\rm SUSY}=1.5~\TeV$. We evaluate
the $\mrh[i]$'s at $\Lambda_L$, and we neglect any possible
running of the $\mD[i]$'s and $\mrh[i]$'s. The so obtained
$\mrh[i]$'s clearly correspond to the scale $\Lambda_L$.


\begin{figure}[!t]\vspace*{-.12in}
\hspace*{-.19in}
\begin{minipage}{8in}
\epsfig{file=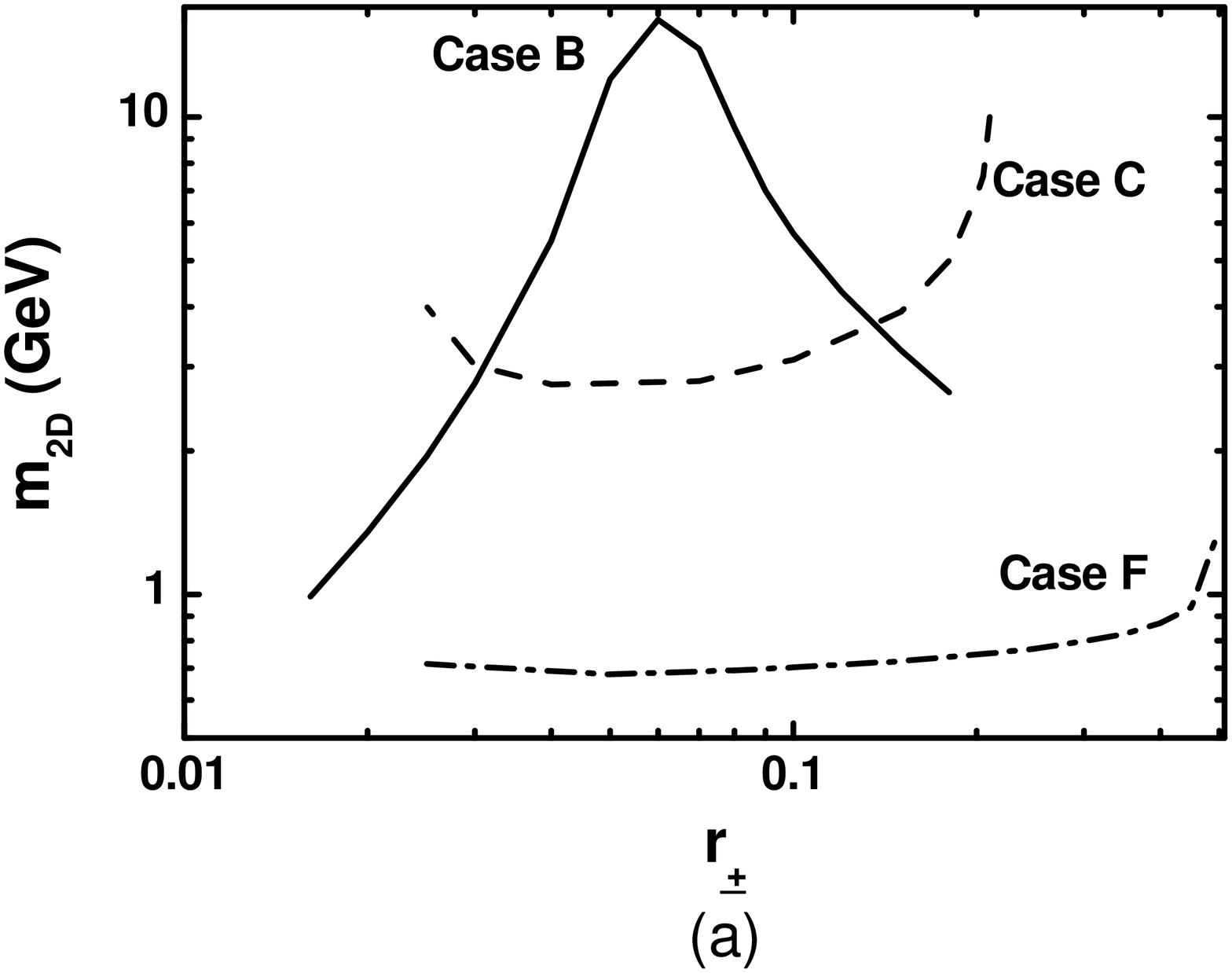,height=3.6in,angle=-90}
\hspace*{-1.2cm}
\epsfig{file=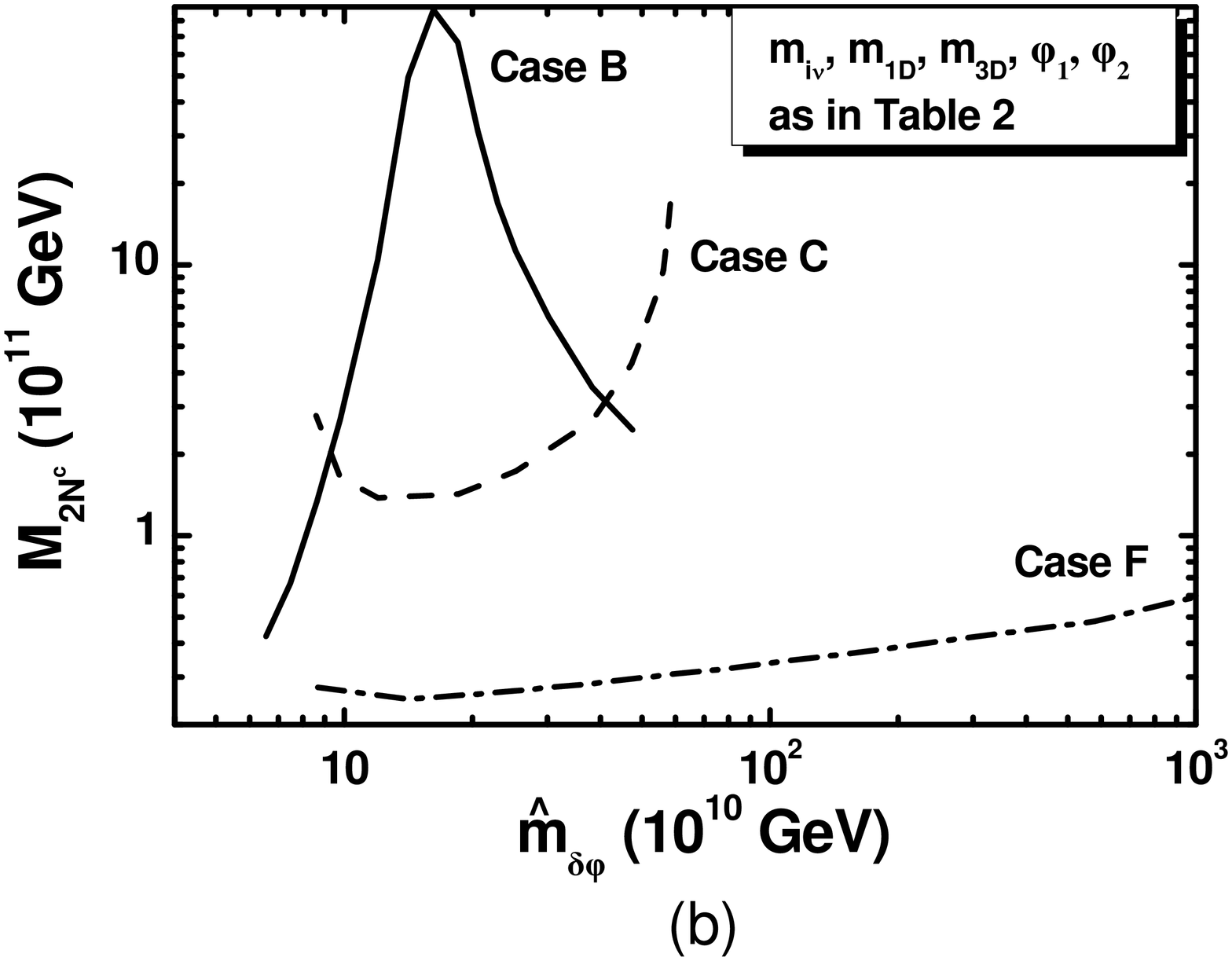,height=3.6in,angle=-90} \hfill
\end{minipage}
\hfill \caption{\sl\small  Contours, yielding the central $Y_B$ in
Eq.~(5.8) consistently with the inflationary requirements, in the
{\sffamily\ftn (a)} $\rs-m_{\rm 2D}$ and {\sffamily\ftn (b)}
$\widehat{m}_{\delta\phi}-M_{2N^c}$ plane. We take $K=K_2$ or
$K_3$ with $\nsu=2$, $\lm=10^{-6}$ and the values of $m_{i\nu}$,
$m_{\rm 1D}$, $m_{\rm 3D}$, $\varphi_1$, and $\varphi_2$ which
correspond to the cases B (solid line), C (dashed line) and F
(dot-dashed line) of Table~4.}\label{fmD}
\end{figure}

Some representative values of the parameters which yield $\Yb$ and
$\Yg$ compatible with \eqs{BAUwmap}{Ygw}, respectively are
arranged in \Tref{tab2}. We take the best-fit $\rs$ value in
\Eref{res2} and $\lm=10^{-6}$ in accordance with \eqs{lm1}{lm2}
with $\nb=2$. We obtain $\msn=8.9\cdot10^{10}~\GeV$ for $K=K_1$
and $\msn=8.6\cdot10^{10}~\GeV$ for $K=K_2$ or $K_3$. Although
such an uncertainty from the choice of $K$'s do not cause any
essential alteration of the final outputs, we mention just for
definiteness that we take $K=K_2$ or $K_3$ throughout. We consider
NO (cases A and B), almost degenerate (cases C, D and E) and IO
(cases F and G) $\mn[i]$'s. In all cases, the current limit in
\Eref{sumnu} is safely met. The gauge symmetry considered here
does not predict any particular Yukawa unification pattern and so,
the $\mD[i]$'s are free parameters. This fact facilitates the
fulfilment of \sEref{kin}{b} since $mD[1]$ affects heavily
$mrh[1]$. Care is also taken so that the perturbativity of
$\ld_{iN^c}$ holds, i.e., $\ld_{iN^c}^2/4\pi\leq1$. The inflaton
$\dphi$ decays mostly into $N_1^c$'s -- see cases A -- E. In all
cases $\GNsn<\Ghsn$ and so the ratios $\GNsn/\Gsn$ introduce a
considerable reduction in the derivation of $\Yb$. In \Tref{tab2}
we also display the values of $\Trh$, the majority of which are
close to $3\cdot10^7~\GeV$, and the corresponding $\Yg$'s, which
are consistent with \Eref{Ygw} for $\mgr\gtrsim1~\TeV$. These
values are in nice agreement with the ones needed for the solution
of the $\mu$ problem of MSSM -- see, e.g., \Tref{tab}.


In order to investigate the robustness of the conclusions inferred
from \Tref{tab2}, we examine also how the central value of $Y_B$
in \Eref{BAUwmap} can be achieved by varying $\rs$ (or $\msn$) and
adjusting conveniently $\mD[2]$ (or $\mrh[2]$) -- see
\sFref{fmD}{a} or {\sf\ftn (b)} respectively. We fix again
$\lm=10^{-6}$. Since the range of $Y_B$ in \Eref{BAUwmap} is very
narrow, the $95\%$ c.l. width of these contours is negligible. The
convention adopted for the various lines is depicted in the plot
of \sFref{fmD}{b}. In particular, we use solid, dashed and
dot-dashed line when the remaining inputs -- i.e. $\mn[i]$,
$\mD[1]$, $\mD[3]$, $\varphi_1$, and $\varphi_2$ -- correspond to
the cases B, C and F of \Tref{tab2}, respectively. Only some
segments from the $\rs$ range in \Eref{res1} fulfill the
post-inflationary requirements. Namely, as inferred by
\sFref{fmD}{a}, we find that $\rs$ may vary in the ranges
$(0.0161-0.18)$, $(0.025-0.21)$ and $(0.025-0.499)$ for $\mD[2]$
plotted in \sFref{fmD}{a} and the remaining inputs of the cases B,
C and F respectively. As regards the other quantities, in all we
obtain
\beq 4.4\lesssim Y_{\Gr}/10^{-15}\lesssim228\>\>\>\mbox{and}\>\>\>
0.23\lesssim\Trh/{10^{8}\GeV}\lesssim12\>\>\>\mbox{with}\>\>\>6.5\lesssim\msn/10^{10}\lesssim4241\,.\label{res3}\eeq
As a bottom line, nTL not only is a realistic possibility within
our setting but also it can be comfortably reconciled with the
$\Gr$ constraint even for $\mgr\sim1~\TeV$ as deduced from
\eqs{res3}{Ygw}.

\section{Conclusions}\label{con}

We investigated the realization of kinetically modified
non-minimal HI and nTL in the framework of a $B-L$ extension of
MSSM endowed with the condition that the GUT scale is determined
by the running of the three gauge coupling constants. Our setup is
tied to the super-{} and \Kap s given in Eqs.~(\ref{Whi}) and
(\ref{K1}) -- (\ref{K3}). Prominent in this setting is the role of
a softly broken shift-symmetry whose violation is parameterized by
the quantity $\rs=\cp/\cm$ and can be constrained by the
observations. Our models exhibit the following features: {\sf\ftn
(i)} they inflate away cosmological defects; {\sf\ftn (ii)} they
safely accommodates observable gravitational waves with
subplanckian inflaton values and without causing any problem with
the validity of the effective theory; {\sf\ftn (iii)} they offer a
nice solution to the $\mu$ problem of MSSM, provided that
$\ld_\mu$ is somehow small; {\sf\ftn (iv)} they allow for
baryogenesis via nTL compatible with $\Gr$ constraints and
neutrino data. In particular, we may have $\mgr\sim1~\TeV$, with
the inflaton decaying mainly to $N^c_1$ and $N^c_2$ -- we obtain
$\mrh[i]$ in the range $(10^{9}-10^{14})~\GeV$. It remains to
introduce a consistent soft SUSY breaking sector in the theory
which is certainly an important and difficult task.


\def\ijmp#1#2#3{{\emph{Int. Jour. Mod. Phys.}}
{\bf #1},~#3~(#2)}
\def\plb#1#2#3{{\emph{Phys. Lett.  B }}{\bf #1},~#3~(#2)}
\def\zpc#1#2#3{{Z. Phys. C }{\bf #1},~#3~(#2)}
\def\prl#1#2#3{{\emph{Phys. Rev. Lett.} }
{\bf #1},~#3~(#2)}
\def\rmp#1#2#3{{Rev. Mod. Phys.}
{\bf #1},~#3~(#2)}
\def\prep#1#2#3{\emph{Phys. Rep. }{\bf #1},~#3~(#2)}
\def\prd#1#2#3{{\emph{Phys. Rev.  D }}{\bf #1},~#3~(#2)}
\def\npb#1#2#3{{\emph{Nucl. Phys.} }{\bf B#1},~#3~(#2)}
\def\npps#1#2#3{{Nucl. Phys. B (Proc. Sup.)}
{\bf #1},~#3~(#2)}
\def\mpl#1#2#3{{Mod. Phys. Lett.}
{\bf #1},~#3~(#2)}
\def\arnps#1#2#3{{Annu. Rev. Nucl. Part. Sci.}
{\bf #1},~#3~(#2)}
\def\sjnp#1#2#3{{Sov. J. Nucl. Phys.}
{\bf #1},~#3~(#2)}
\def\jetp#1#2#3{{JETP Lett. }{\bf #1},~#3~(#2)}
\def\app#1#2#3{{Acta Phys. Polon.}
{\bf #1},~#3~(#2)}
\def\rnc#1#2#3{{Riv. Nuovo Cim.}
{\bf #1},~#3~(#2)}
\def\ap#1#2#3{{Ann. Phys. }{\bf #1},~#3~(#2)}
\def\ptp#1#2#3{{Prog. Theor. Phys.}
{\bf #1},~#3~(#2)}
\def\apjl#1#2#3{{Astrophys. J. Lett.}
{\bf #1},~#3~(#2)}
\def\n#1#2#3{{Nature }{\bf #1},~#3~(#2)}
\def\apj#1#2#3{{Astrophys. J.}
{\bf #1},~#3~(#2)}
\def\anj#1#2#3{{Astron. J. }{\bf #1},~#3~(#2)}
\def\mnras#1#2#3{{MNRAS }{\bf #1},~#3~(#2)}
\def\grg#1#2#3{{Gen. Rel. Grav.}
{\bf #1},~#3~(#2)}
\def\s#1#2#3{{Science }{\bf #1},~#3~(#2)}
\def\baas#1#2#3{{Bull. Am. Astron. Soc.}
{\bf #1},~#3~(#2)}
\def\ibid#1#2#3{{\it ibid. }{\bf #1},~#3~(#2)}
\def\cpc#1#2#3{{Comput. Phys. Commun.}
{\bf #1},~#3~(#2)}
\def\astp#1#2#3{{Astropart. Phys.}
{\bf #1},~#3~(#2)}
\def\epjc#1#2#3{{Eur. Phys. J. C}
{\bf #1},~#3~(#2)}
\def\nima#1#2#3{{Nucl. Instrum. Meth. A}
{\bf #1},~#3~(#2)}
\def\jhep#1#2#3{{\emph{JHEP} }
{\bf #1},~#3~(#2)}
\def\jcap#1#2#3{{\emph{JCAP} }
{\bf #1},~#3~(#2)}
\def\jcapn#1#2#3#4{{\sl JCAP }{\bf #1}, no. #4, #3 (#2)}
\def\prdn#1#2#3#4{{\sl Phys. Rev. D }{\bf #1}, no. #4, #3 (#2)}
\newcommand{\arxiv}[1]{{\ftn\tt  arXiv:#1}}
\newcommand{\hepph}[1]{{\ftn\tt  hep-ph/#1}}
\def\prdn#1#2#3#4{{\sl Phys. Rev. D }{\bf #1}, no. #4, #3 (#2)}
\def\jcapn#1#2#3#4{{\sl J. Cosmol. Astropart.
Phys. }{\bf #1}, no. #4, #3 (#2)}
\def\epjcn#1#2#3#4{{\sl Eur. Phys. J. C }{\bf #1}, no. #4, #3 (#2)}

\end{document}